\documentclass[
 twocolumn,
]{ceurart}
\usepackage{xspace}
\usepackage{listings}
\begin{document}

\copyrightyear{2022}
\copyrightclause{Copyright for this paper by its authors. Use permitted under Creative Commons License Attribution 4.0 International (CC BY 4.0).}

\conference{4th Edition of Knowledge-aware and Conversational Recommender Systems (KaRS) Workshop @ RecSys 2022, September 18--23 2023, Seattle, WA, USA.}
\newcommand{\infact}{\textsc{INFACT}\xspace}
\title{INFACT: An Online Human Evaluation Framework for Conversational Recommendation}

\author{Ahtsham Manzoor}[%
orcid=0000-0001-9418-753,
email=ahtsham.manzoor@aau.at,
url=https://ahtsham58.github.io/,
]
\cormark[1]
\address{University of Klagenfurt, Universitätsstraße 65-67, Klagenfurt am Wörthersee, 9020, Austria }

\author{Dietmar Jannach}[%
orcid=0000-0002-4698-8507,
email=dietmar.jannach@aau.at,
url=https://www.aau.at/en/aics/research-groups/infsys/team/dietmar-jannach/,
]

\cortext[1]{Corresponding author.}

% ======================
% REVIEW TODOs
% - I suggest that the authors add some details regarding the difference between the data collected via the INFACT framework and the previous efforts in the literature. How does the concept of "meaningfulness" differ from the attributes collected in other studies (e.g., [9, 14])?
%- Training crowdworkers is a challenging task, especially for tasks that are strongly subjective, as in NLP. Therefore, I think that the authors should dedicate more space to the training of the human evaluators.
%- In Section 3, the authors present different ways to distribute the workload among the evaluators. Since the authors have already done experiments using the framework, it would be helpful to provide a brief insight of the pros and cons of each option. In fact, it would be useful to include a new section dedicated to the intuitions collected while deploying the INFACT framework.

\begin{abstract}
Conversational recommender systems (CRS) are interactive agents that support their users in recommendation-related goals through multi-turn conversations. Generally, a CRS can be evaluated in various dimensions. Today's CRS mainly rely on \emph{offline} (computational) measures to assess the performance of their algorithms in comparison to different baselines. However, offline measures can have limitations, for example, when the metrics for comparing a newly generated response with a ground truth do not correlate with human perceptions, because various alternative generated responses might be suitable too in a given dialog situation. Current research on machine learning-based CRS models therefore acknowledges the importance of humans in the evaluation process, knowing that pure offline measures may not be sufficient in evaluating a highly interactive system like a CRS.

In this work, we provide a user-centric evaluation approach to conversational recommendation along with the \textbf{\infact}, an onl\textbf{I}ne huma\textbf{N} evaluation \textbf{F}ramework for convers\textbf{A}tional re\textbf{C}ommender sys\textbf{T}ems,
which can be used to assess the suitability of system responses in a given dialog situation. The \infact framework is prepared to enable the crowdsourcing of the evaluation task, where various CRS   can be integrated for comparison. We have successfully applied the \infact framework for conducting a number of user studies in our previous research. We believe that our study design along with the \infact framework can be helpful in facilitating user-centric studies in domains such as dialog systems, machine translation, or Q\&A. We release the source code of the framework at \textcolor{blue}{\url{https://github.com/ahtsham58/INFACT}}.

\end{abstract}


\maketitle

\section{Introduction}

Conversational recommender systems (CRS) support their users in finding items of interest through multi-dialogs, often in natural language \cite{jannach2021crscsur}. A CRS is generally considered a highly interactive system, where users converse with the agent and seek for recommendations. Due to the highly interactive nature, modern CRS are generally complex and consist of multiple components, see e.g., \cite{jannach2021crscsur,chen2021knowledge, ArbpitNavigationByPreference2020,Jannach:2004:ASK:3000001.3000153}. Overall, the eventual goal of a CRS is to support non-trivial yet useful interactions with their users \cite{jannach2022aimagcrs}.

Evaluating the usefulness of CRS in the academic environment is generally challenging and can be both time and resource intensive in particular when humans are involved in the loop, see also \cite{jannach2022evaluatingcrs}.  For example, assessing the quality of responses and thereby dialogs is as important as assessing the quality of the underlying \emph{recommendation algorithms}. Mostly the research community relies on \emph{offline} evaluation approaches using historical datasets in order to understand how good an algorithm performs. Such an approach can be appropriate in evaluating the prediction capability of an algorithm, e.g., which item a user will consider to consume or rate highly. However, these evaluation approaches are unable to inform about the quality perceptions from a user's perspective. For example, whether the made recommendations are acceptable to the user or if the made recommendations will assist users in discovering new yet relevant items.

Furthermore, assessing linguistic aspects such as consistency, naturalness or fluency as a proxy for the language quality of the system's responses is challenging in its own. In this context, to assess language quality, researchers mainly apply offline metrics, e.g., distinct N-gram and Perplexity, or they compare the system's responses with their ground truths using the BLEU \cite{Papineni2002BLEU} or NIST \cite{doddington2002automatic} scores, see, e.g., \cite{nie2019multimodal, chen2019towards}. However, these offline metrics do not inform us whether the response is grammatically and semantically complete or if the response is \emph{meaningful} given the previous dialog history. Moreover, in reality, the system might respond to a user's utterance in a meaningful way, but may not match the ground truth \cite{manzoor2021towardscrbcrs}. Ultimately, offline linguistic metrics may therefore not fully inform us about the users' quality perceptions in practice.

Current research in CRS acknowledges the importance of humans in the evaluation process, and this also holds for most recent ``end-to-end'' learning approaches, where deep neural network models are trained using recommendation dialogs collected between humans, see, e.g., \cite{chen-etal-2019-towards, zhou2020improving, zhou2021crfr, zhou2020towards, zou2022improving}. In these recent works, we therefore find studies involving humans, and experiments are conducted using various evaluation methodologies.  However, such evaluations often have limitations.
For example, in \cite{li2018towards} human judges were asked to provide a \emph{relative} ranking of the responses by different systems. In case of relative comparison, it remains unfortunately unclear if any of the compared systems are useful at all \cite{JannachManzoor2020}. Moreover, the scope of such studies seems limited as in many cases there are only a few evaluators involved and sometimes the details regarding the background of the human judges are missing too. Also, often only the language quality of the responses is the main focus of such studies. An assessment if the made recommendations are suitable in an ongoing dialog context is sometimes missing, see, e.g., \cite{zhou2021crfr,hayati2020inspired}.

In this work, we present a user-centric evaluation approach to CRS that can be used to assess both linguistic and recommendation quality aspects along with the \textbf{\infact}, an onl\textbf{I}ne huma\textbf{N} evaluation \textbf{F}ramework for convers\textbf{A}tional re\textbf{C}ommender sys\textbf{T}ems.
We have applied our evaluation approach for a number of studies \cite{manzoor2021towardscrbcrs,MANZOOR2021100139,manzoorrs2021}, using the \infact framework as a basis.  To easily involve a larger set of subjects than in earlier studies, the \infact framework % , where different systems can be integrated for comparison,
is prepared to support the evaluation task through online crowdworking platforms. We believe that our study design may serve as a blueprint for future human evaluation studies for CRS. It can furthermore be easily extended to evaluate dialogs systems, machine translation, or Q\&A tasks. We release the source code of the \infact framework at \url{https://github.com/ahtsham58/INFACT}.

\section{Related Work}
According to recent surveys on CRS \cite{jannach2021crscsur,jannach2022evaluatingcrs}, we can generally observe three main dimensions in which a CRS can be evaluated: \emph{(i) effectiveness of task support}, i.e., the ability of the system to support a recommendation-related task, \emph{(ii) efficiency of task support}, i.e., how much effort is required by the user, and \emph{(iii) conversation quality and usability}, which may cover aspects like fluency, naturalness, or the consistency of the system responses. All of these aspects can contribute to the success of a CRS in practice.

From a methodological standpoint,  quality measurements are typically either made with the help of computational (``offline'') experiments or with studies involving humans in the loop. In offline experiments, system effectiveness  is often evaluated in terms of \emph{recommendation quality}, where   metrics like \emph{precision} or \emph{recall} are used as proxies. In addition to recommendation quality, offline experiments are common for assessing the \emph{dialog quality}. Specifically, linguistic measures such as \emph{distinct N-gram} and \emph{perplexity} to assess the diversity and fluency of the system-generated responses have been applied in various recent works on CRS, see, e.g., \cite{chen2021knowledge, chen-etal-2019-towards, zhou2020improving, zhou2021crfr}. Similarly, inspired by the machine translation domain,   metrics like \emph{BLEU} or \emph{NIST} are applied in several works on CRS, where the system response is compared with a given ground truth in order to estimate the overall quality of the generated responses, see, e.g., \cite{chen2021knowledge, zhou2020towards, hayati2020inspired, 8260795}.

Given the interactive nature of CRS, studies involving humans in the evaluation process are not uncommon in the CRS literature. Such studies mainly assess the quality aspects from a user's perspective. See \cite{Pu:2011:UEF:2043932.2043962,jin2021key,jannach2022evaluatingcrs} for a set of relevant quality attributes. Looking at most recent works, different studies were conducted in which human judges were tasked to \emph{rate} or \emph{rank} system responses in various dimensions.  For example, in \cite{li2018towards}, ten evaluators were asked to rank the responses by baseline recommenders and the proposed system in terms of the overall quality. Similarly, in \cite{hayati2020inspired}, the authors reported a study in which human evaluators had to rate the responses on a scale from 1-5 in terms of \emph{Fluency}, \emph{Consistency},
\emph{Naturalness}, \emph{Persuasiveness}, and \emph{Engagingness}. Similarly, in a recent work \cite{zhou2021crfr}, five human judges were given the task to rate the responses generated by the proposed system and various baselines on a scale from 1-3 in terms of  \emph{Fluency}, \emph{Coherence}, \emph{Informativeness}, and \emph{Interoperability}. Similar examples of such studies can also be found in \cite{zhou2020improving,chen-etal-2019-towards,zhou2020towards,Pecune:HAI2019}.

Interestingly, such studies mainly focused on \emph{linguistic} aspects of the systems' responses. If the recommendations themselves were considered meaningful---a main aspect in terms of a system's usefulness---was not assessed with the help of human judges but rather evaluated through offline analyses. On the other hand, the generic concept of \emph{meaningfulness} of a response can be used to evaluate both aspects, i.e., language and recommendation quality, see, e.g., \cite{manzoor2021towardscrbcrs, MANZOOR2021100139, manzoorrs2021}.
Moreover, in such studies the details about the study setup and background of the evaluators were quite brief and sometimes missing at all. Furthermore, several studies were conducted with a small  number of judges, for example, in \cite{chen-etal-2019-towards,zhou2020improving}, \emph{three} judges were involved and no information was provided regarding their linguistic expertise.
%In addition, from a methodological point of view, human evaluation approaches where \emph{relative ranking} of systems was adapted, as was done in \cite{li2018towards}, has some limitations. For example, in case of relative comparison, it is unclear if any of the compared systems are useful at all \cite{JannachManzoor2020}.

Given the potential limitations of offline experiments and of user studies with unclear significance, we provide a study design that may serve as a template for scalable human-centric evaluation studies of dialog systems. Next, we explain the experiment design of our evaluation approach and highlight the features that the \infact framework offers to support the evaluation task through online crowdworking platforms. %Finally, we conclude the paper with a discussion on the implications of our research.

\begin{figure*}[h!t]
  \centering
  \includegraphics[width=1.0\textwidth]{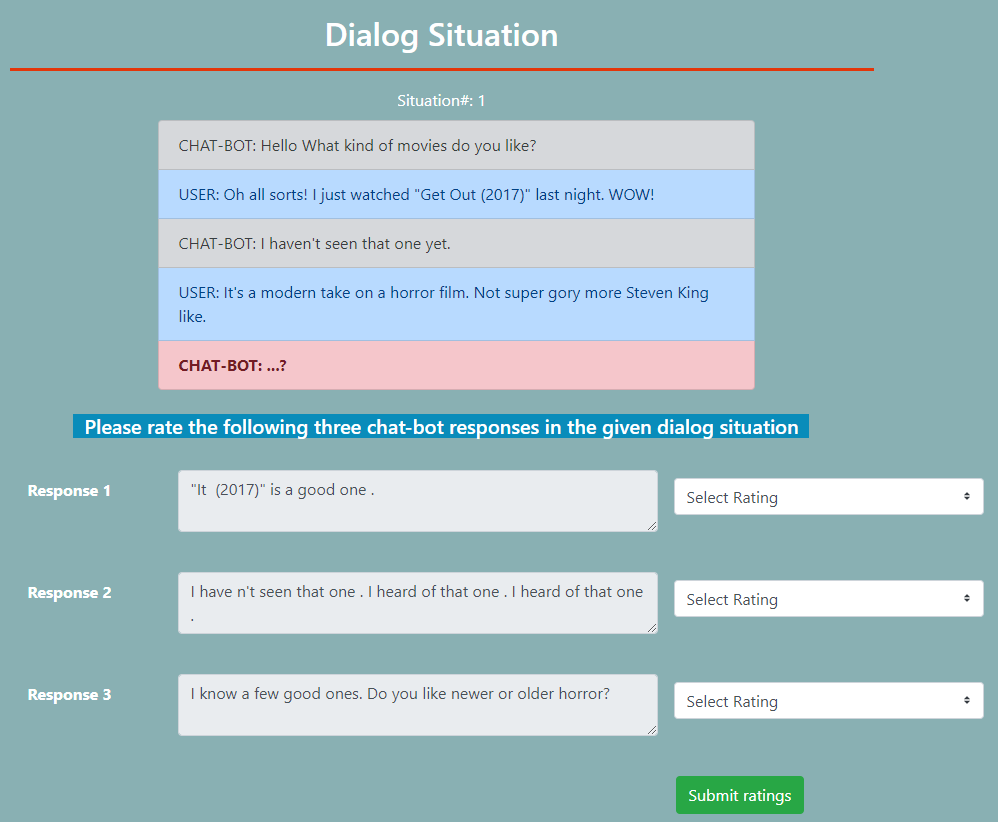}
 \caption{Response rating user interface}
  \label{fig:dialog}
\end{figure*}

\begin{comment}
\begin{figure*}[h!t]
  \centering
  \includegraphics[width=1.0\textwidth]{fig-infect-archit2.png}
 \caption{\infact System Architecture}
  \label{fig:archi}
\end{figure*}
\end{comment}

\section{Experiment Design with \infact}
\paragraph{General Design}
The \infact framework is developed based on the concept that user-centric evaluations are vital to assess the effectiveness of highly interactive systems like CRS. Moreover, evaluating such systems requires studies at scale in order to investigate the quality of both linguistic and recommendation aspects in practice. Specifically, in our approach we ask human subjects to assess dialog continuations (``system responses'') provided by a CRS given a piece of dialog (``dialog situation'') using one or more quality criteria. In our own studies, we use the ReDial dataset \cite{li2018towards} consisting of real-world dialogs for such evaluations. We note that various recent current CRS approaches rely on this dataset or similar ones to \emph{generate} or \emph{retrieve} suitable responses, e.g., \cite{zhou2020towards,hayati2020inspired,kang2019recommendation}.  To avoid biases and to receive feedback for all stages of the dialog, the dialog situations to be evaluated are selected from such datasets at random \cite{cai2019department,lyu2021workflow}. In addition, we assume to have a larger set of participants than many existing studies; in particular we consider crowdsourcing to be helpful.

Such an approach can be instantiated in various ways, depending on the research question(s). For example, an experiment could include one or more algorithms to evaluate for each participant. Similarly, there could be one or more questions regarding the dialog quality and for each dialog continuation. Moreover, each participant can be tasked to assess only one or more dialog situations and the feedback scale could be different too.

When deciding on these specifics, it is important to keep the cognitive load and the overall workload for the study participants in mind. Moreover, the specific design also determines how many participants are required to achieve a sufficient number of human judgements. In our own experiments, as discussed later, we decided to ask participants to assess exactly three different dialog continuations, and they had to assess ten such dialog situations. As a result, we obtain multiple assessments from each participant, which helps to keep the number of participants low. However, using such a design, it is important to check for intraclass (per user) correlations in the statistical analyses.

Arbitrary dialog datasets can be used with our framework, as long as they follow the format used in the framework, which is currently based on the ReDial dataset, as mentioned. Finally, arbitrary post-task questionnaire elements can be introduced, and the \infact framework implements a number of typically required functionalities, e.g., for persistently storing the feedback into a database.

 %Also, various  attention checks, for example study completion time, can be implemented in both dialog continuations or post-task questionnaire to deal with the potentially unreliable crowdworkers.
\paragraph{A Specific Implementation}
In our experiments \cite{manzoor2021towardscrbcrs, MANZOOR2021100139,  manzoorrs2021}, each participant is presented with dialog situations that always start from the first utterance and end with a \emph{user} utterance. Below the dialog situation, as shown in Figure \ref{fig:dialog},  we show three responses to the last user utterance by three different CRS under comparison. To  highlight and differentiate item recommendations from regular language words in the dialog situation and responses, we enclosed item names, e.g., in this case movie titles, in double quotes\footnote{A detailed description about how to prepare the evaluation data is explained in the \textcolor{blue}{ \href{https://github.com/ahtsham58/INFACT}{online}} repository.}.

%However, such a design decision has to be made while preparing evaluation data\footnote{A detailed description about how to prepare the evaluation data is explained in the \textcolor{blue}{ \href{https://github.com/ahtsham58/INFACT}{online}}repository.}.

\begin{figure*}[h!t]
  \centering
  \includegraphics[width=1.0\textwidth]{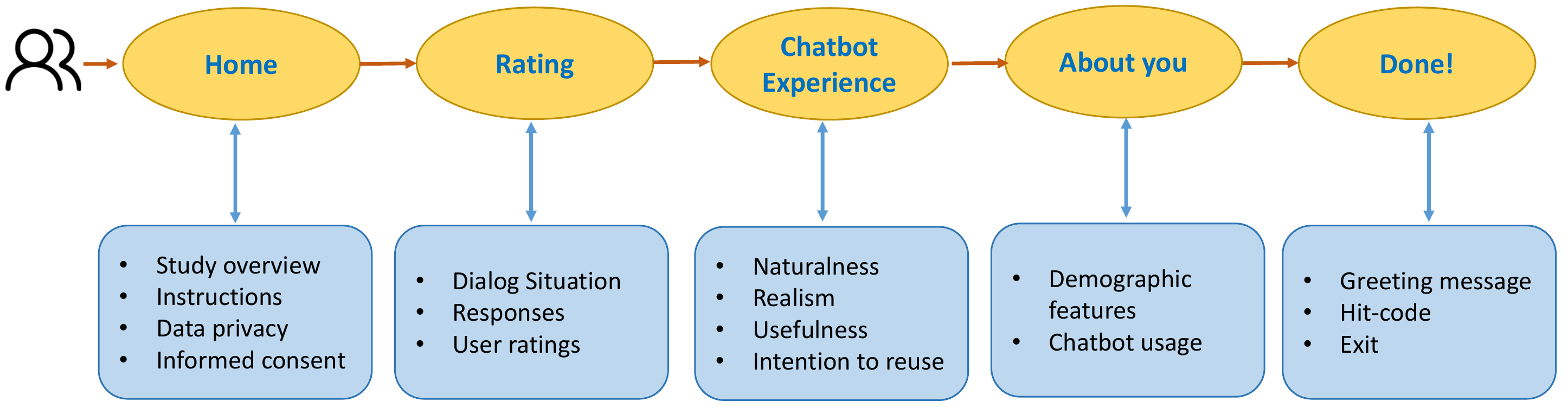}
 \caption{\infact  Workflow Diagram}
  \label{fig:workflow}
\end{figure*}

In our experiments, the only question for the study participants was to \emph{independently} assess (or rate) the quality of each response in terms of the \emph{meaningfulness} of the responses in the given dialog context. To obtain fine-grained assessments, we use a 5-point scale labeled from `Entirely meaningless' to `Perfectly meaningful', which can be modified depending on the research question. To avoid any sequential rating bias by the evaluators, the order of showing responses to the user is randomized.

On the landing page, we provide specific instructions to the evaluators about how to judge the \emph{meaningfulness} of responses. For example, a response by the specific system should be logical continuation of the provided dialog situation. In case an item recommendation(s) is included in the response, it has to match to the user's stated interest and preferences. If the system response does not include a movie recommendation, e.g., chit-chat sentence, the participants are supposed to rate the \emph{meaningfulness} of the response as a reply to the user's last utterance while considering also the context of the ongoing dialog. Overall,  human judges are supposed to provide ratings based on their \emph{subjective} quality perceptions.

From the linguistic perspective, a deep discussion of the concept `\emph{meaningfulness}' is provided in \cite{10.2307/2252520}, where the author makes a distinction between `grammaticality' and `meaningfulness'. In our study design, instead of challenging participants with complex linguistic concepts or considerations, we provide  examples and guidelines when a response can be considered meaningful or meaningless.\footnote{The term ``meaningfulness'' is also used in the context of a human evaluation in \cite{nie2019multimodal}. Differently from our work, the term ``meaningful'' is used in \cite{nie2019multimodal} to summarize other evaluation dimensions in an informal way. We note that our framework can easily be configured to collect annotator feedback on several dimensions, e.g., fluency, coherence, or informativeness as in \cite{zhou2021crfr}.}

%For further details on the concept of \emph{meaningfulness} in the context of CRS, see also \cite{manzoor2021towardscrbcrs}.

In this way, each evaluator assessed \emph{ten} such dialog situations. However, this is a configurable parameter that can changed based on the experiment design. On submission, we store the rating scores, including the dialog situation, corresponding responses, and the overall time it took for the evaluator to rate, in a NoSQL cloud database. %\footnote{\url{https://firebase.google.com/}}.
We relied on a NoSQL database as it offers flexible, affordable, and scalable database management.

\begin{table}[h!t]
\caption{Example Questions from a Post-Task Questionnaire}
  \centering
  \begin{tabular}{lp{6cm}}
  \hline
    \textbf{} & \textbf{Questions}  \\ \hline
   Q1 & I found the presented dialogues natural. \\
 Q2 & The presented dialogue situations look realistic. \\
  Q3 & I could imagine that such dialogues also happen between humans. \\
 Q4 & Considering only the best responses found in each dialogue, I would find the chat-bot useful. \\
 Q5 & Considering only the best responses found in each dialogue, I would probably use such a movie recommendation chat-bot in the future. \\
  \hline
\end{tabular}
\label{tab:questionnaire-items}
\end{table}
\begin{table} [h!t]
\caption{Example questions for Participant Demographics}
\label{tab:demographics}

\renewcommand\thetable{1}
  \centering
%\resizebox{\textwidth}{!}{
\small
\begin{tabular}{p{3.7cm}c}
\hline
 \textbf{Demographic Feature} & \textbf{Scale} \\
\hline
\multirow{3}{4.2cm}{Gender}	
                  & Male  \\
                  & Female \\
                  & Other
                  \\\hline
\multirow{5}{4.2cm}{Age}
                  & 18-25  \\
                  & 25-30 \\
                  & 30-35 \\
                  & 35-45 \\
                  & 45-70
                   \\\hline
\multirow{4}{4.2cm}{English fluency level}
                      & Beginner  \\
                      & Intermediate \\
                      & Fluent \\
                      & Advanced
                      \\\hline

\multirow{5}{4.2cm}{Education level}
                      & High school or less \\
                      & Bachelor's \\
                      & Master's \\
                      & Doctorate \\
                      & Other
                      \\\hline
\multirow{5}{4.2cm}{Frequency of watching movies}
                      & Everyday  \\
                      & Several times a week  \\
                      & Once in a week \\
                      & Once every few weeks \\
                      & Less frequent
                      \\\hline
\multirow{2}{4.2cm}{Ever interacted with a chat-bot}
                      & Yes  \\
                      & No
                      \\\hline
\multirow{2}{4.2cm}{Ever interacted with a chat-bot for getting movie recommendations}
                      & Yes  \\
                      & No
                      \\\hline

\end{tabular}
\normalsize
%} % resizebox
\end{table}

In order to check if participants are attentive during the study, one of the ten dialog situations that appear in random order is used as an \emph{attention check}. Specifically, in one of the three responses for this dialog situation, we asked the participants to select a particular rating from the given scale. The attention check was considered to be failed whenever a study participant did not select the required score. In this case, we completely discard all data from such unreliable crowdworkers\footnote{We provide the Python script to automatically parse the study data stored in the JSON format on the cloud.}. Furthermore, apart from the explicit attention check, the \infact framework is equipped with various \emph{implicit} checks to deal with the potentially unreliable crowdworkers such as time interval (in seconds) for each individual event, overall study completion time, etc.

In our specific experiments, no particular training or expertise is required by the crowdworkers to participate. To fulfill the task, the crowdworkers were asked for their subjective assessment regarding the generated system responses in terms of their meaningfulness. In case participants should answer more complex questions, e.g., regarding fluency or interpretability as in \cite{zhou2021crfr}, appropriate measures must be taken to ensure that the crowdworkers are able to fulfill the task reliably, e.g., by providing more instructional material or by requiring certain skills.

After the submission of ratings for ten such dialog situations, a post-task questionnaire is shown to the participants, where we collect general feedback regarding the quality of dialogs, demographics, and general remarks or suggestions. Table \ref{tab:questionnaire-items} shows parts of a questionnaire for dialog quality, asking, for example, if the shown dialogs feel natural, realistic, or useful. An example list of demographic questions is shown in Table \ref{tab:demographics}. These questionnaires can be modified depending on the research question(s). The overall workflow of our experiment design is visualized in Figure \ref{fig:workflow}.

To enable and support studies through crowdworking platforms like Amazon Mechanical Turk, Prolific, etc., we used a pre-implemented feature like hit-code generation. %\footnote{A  demo of the \infact evaluation framework can be seen  at \textcolor{blue}{\url{https://study-ainf.aau.at/}}}.
Our assumption is that a large number of human judges are needed for the evaluation, hence the \infact framework is prepared accordingly. Technically, the \infact framework is a web-based application developed using the Django framework in Python 3.0, and Bootstrap
4.4.

%Finally, to analyze the results, the study data stored on the cloud DB in the JSON format can be parsed using the provided Python script.

\section{Conclusion}
Research on conversational recommender systems (CRS) has attracted increased attention in recent years.  The most recent proposals on CRS, and in particular ones that follow an end-to-end learning paradigm, mainly rely on computational measures in order to demonstrate the effectiveness of their systems in comparison to different baselines. However, the aspects that contribute to the success or failure of a CRS may not be fully assessed without involving humans in the evaluations process.
%Given the interactive nature of CRS, the research community has made substantial progress in devising various accepted evaluation standards. Mainly, such standards are adopted from information retrieval and machine learning domains. The most recent proposals on CRS particularly the ones that follow an end-to-end learning paradigm therefore mainly rely on computational measures in order to demonstrate the effectiveness of their systems in comparison to different baselines. However, the aspects that contribute to the success or failure of a CRS may not be reasonably investigated without involving humans in the evaluations process.

In this work we provide a user-centric evaluation approach for CRS, which can be used to investigate both recommendation and linguistic quality aspects of system responses in a given dialog. Since the scope of several studies reported in the context of recent CRS seems limited, we propose an online evaluation tool which can be used to perform human evaluations at scale with the help of crowdworkers.
%To that end, we believe that the \infact framework can facilitate in crowdsourcing the evaluation task, while conducting studies with the help of substantial number of crowdworkers. Third,
Due to the modular and flexible nature of the architecture underlying  the \infact framework, it can be modified and adapted in the context of a similar study design. For example, replacing the scale or metric requires only a few modifications on a single template page.  Ultimately, we hope that our user-centric evaluation approach can be considered as a template to facilitate the design of similar studies in domains like dialog systems, Q\&A or machine translation.

%% Define the bibliography file to be used
\balance
\bibliography{main}

\end{document}